\documentclass[12pt, draftclsnofoot, onecolumn]{IEEEtran}
\usepackage{graphicx}
\usepackage{epstopdf}
\usepackage[latin1]{inputenc}
\usepackage{amsmath,amssymb,amscd,latexsym,dsfont}
\usepackage[english]{babel}
\usepackage{tabularx,cite}
\usepackage{graphicx}
\usepackage{algpseudocode}
\usepackage{algorithm}
\usepackage{algorithmicx}
\usepackage{float}
\usepackage{multicol}
\usepackage{psfrag}
\usepackage{comment,psfrag,subfigure,enumerate}
\usepackage{color}
\usepackage{amsthm}

\ifCLASSINFOpdf
\else
\fi
\begin{document}
%
\title{Coded-Caching using Adaptive Transmission}

\author{
    \IEEEauthorblockN{Behrooz Makki, ~\IEEEmembership{Senior Member,~IEEE}, Mohamed-Slim Alouini,~\IEEEmembership{Fellow,~IEEE} }
    \thanks{B. Makki is with Ericsson Research, Sweden, Behrooz.makki@ericssom.com. M.-S. Alouini is with the King Abdullah University of Science and Technology (KAUST), Thuwal, Makkah Province, Saudi Arabia, Email: slim.alouini@kaust.edu.sa.}
}
\maketitle

\begin{abstract}
Coded-caching is a promising technique to reduce the peak rate requirement of backhaul links during high traffic periods. In this letter, we study the effect of adaptive transmission on the performance of coded-caching based networks. Particularly, concentrating on the reduction of backhaul peak load during the high traffic periods, we develop adaptive rate and power allocation schemes maximizing the network successful transmission probability, which is defined as the probability of the event with all cache nodes  decoding their intended signals correctly. Moreover, we study the effect of different message decoding and buffering schemes on the system performance. As we show, the performance of coded-caching networks is considerably affected by rate/power allocation as well as the message decoding/buffering schemes.
\end{abstract}
\IEEEpeerreviewmaketitle
\vspace{-4mm}
\section{Introduction}
In 5G and beyond, the number of wireless devices, and their rate requirements, will increase by orders of magnitude \cite{paperG6}. To cope with such demands, different methods are proposed to improve the capacity and spectral efficiency. Here, one of the promising techniques is network densification, i.e., the deployment of many base stations (BSs) of different types such that there are more resource blocks per unit area. The BSs, however, need to be connected to the operators' core network via a transport network, the problem which becomes challenging
as the number of the BSs/users increases. Particularly, the increase of backhaul traffic may lead to backhaul congestion which, in turn, leads to end-to-end latency increment. This is the main motivation for wireless caching schemes reducing the backhaul  load. 

Caching is defined as storing popular reusable information at intermediate nodes to reduce the backhauling load. Such a technique is of interest in delay- and/or backhaul-constrained applications such as D2D, V2X and integrated access and backhaul.

Not every type of information is cacheable, for example, interactive applications such as gaming and voice calls. However, most of the network traffic today, including trending tweets, breaking news and video, is cacheable. Here, particular attention is paid to video. This is because, according to \cite{video2022}, by 2022, $79\%$ of the world's mobile traffic will be video. For instance, Netflix and YouTube alone account for almost half of peak downstream traffic in USA \cite{youtube}. Also, video is typically long and pre-recorded, which makes planning, prediction and segmentation easy. 
Specially, as demonstrated in Fig. 1, video has high variation of the daily traffic profile, and increases the backhaul peak rate during the night significantly. This is important because the wireless network is designed based on the peak traffic. Then, with a variant daily traffic profile, the network will be underutilized most of the time, which is not economically viable. Thus, predicting the videos of interest in the high-traffic (HT) periods and caching them in the access points close to the devices during the low-traffic (LT) periods will lead to considerable cost reduction. For these reasons, caching is currently used by different content providers where, for instance, caches serve approximately $60\%-80\%$ (resp. up to $90\%$) of the Netflix \cite{Netflix2} (resp. Facebook \cite{facebook}) content requests.

\begin{figure}
\vspace{-2mm}
\centering
 \includegraphics[width=0.6\columnwidth]{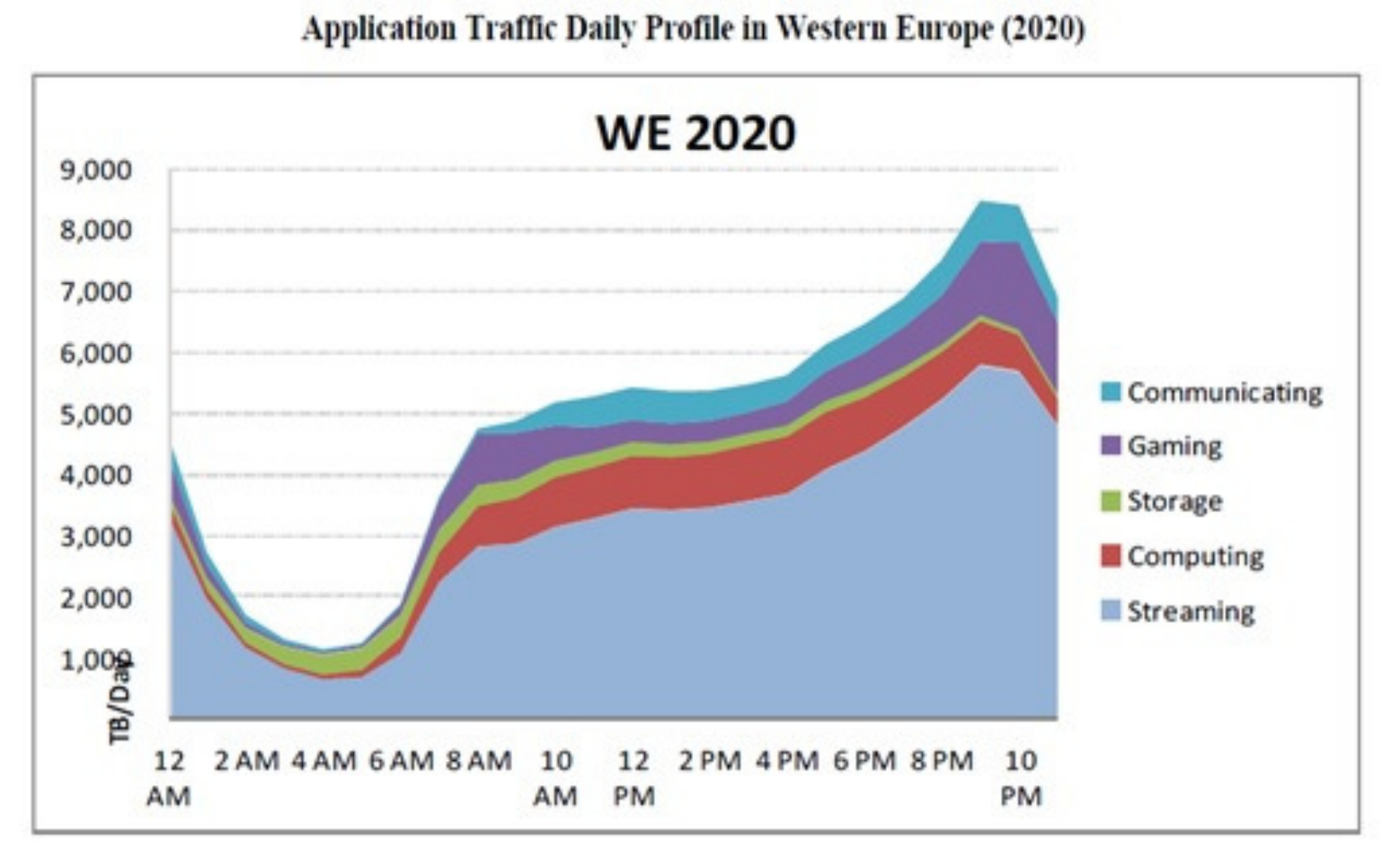}\vspace{-2mm}
\caption{Application traffic daily profile in western Europe \cite[Fig. A3.3]{ssitu}.}\label{figure111}
\vspace{-4mm}
\end{figure}
The initial (un-coded) caching schemes where based on distributing the same information between the cache nodes, and minimizing the cache miss probability defined as the probability of the event that the device's requested file is not in the cache node. Here, depending on the amount of the information available about the sequence of the future requested files, different, e.g., Belady, highest-popularity-first, least-frequently-used and least-recently-used, algorithms have been proposed \cite{7080842}. Minimizing the cache miss probability improves the average system performance. On the other hand, the fundamental work of M. A. Maddah-ali and U. Niesen \cite{6763007,6807823} exploited the multicasting opportunity of cache networks and network coding concept to introduce coded-cashing, minimizing the worst-case backhaul peak rate in HT period. Also, following \cite{6763007,6807823}, there have been several works on the performance analysis of coded-caching systems, e.g., \cite{7857805,8674819,cairecodedcaching}. In these works, it is mainly concentrated on proper partitioning and distribution of the sub-packets between the cache nodes while the wireless channel between the serve and the cache nodes has been rarely studied, e.g. \cite{9014575}. 

In this letter, we study the effect of adaptive data transmission and different data decoding/buffering schemes on the performance of wireless coded-caching networks. We present adaptive rate and power allocation schemes between the sub-packets such that the network successful transmission probability (STP) is maximized. Here, STP is defined as the probability of the event that all cache nodes can decode their intended signals correctly. Moreover, we investigate the effect of different decoding and buffering schemes on the network STP. We concentrate on the worst-case peak backhaul traffic cases where the cache nodes request for different signals during the HT periods. As we show, the performance of coded-caching networks is considerably affected by the decoding scheme as well as rate/power allocation. 

%
%
\section{System Model}

\begin{figure}
\vspace{-2mm}
\centering
 \includegraphics[width=0.6\columnwidth]{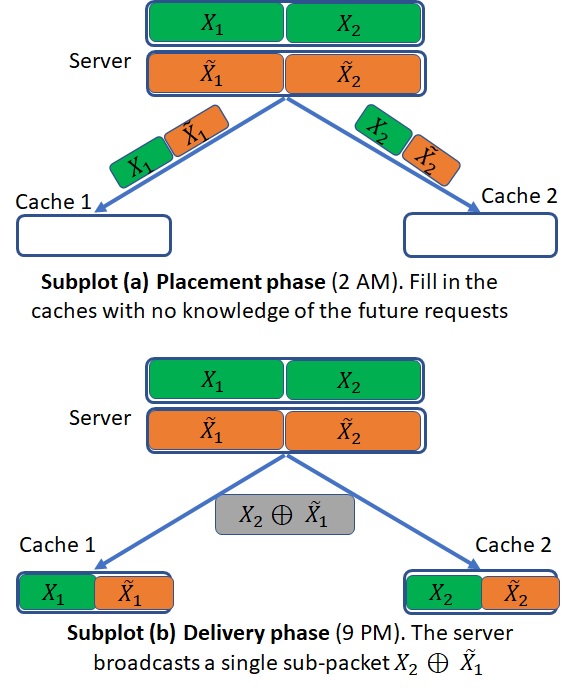}\vspace{-2mm}
\caption{Coded-caching concept. Subplot A: Placements phase (at 2 AM) where the cache nodes are filled in with different sub-packets. Subplot B: Delivery phase (at 9 PM) where a single subpacket is braodcasted to the cache nodes. }\label{figure222}
\vspace{-5mm}
\end{figure}
As illustrated in Fig. 2, consider the simplest case of coded-caching networks where a server connects to two cache nodes $\text{C}_1$ and $\text{C}_2$. However, it is straightforward to extend the results to the cases with different numbers of cache nodes. In general, coded-caching has two, namely, placement and delivery, phases. During the placement phase, performed in the LT period (say, at 2 AM), the server divides the packets, for instance, packets $X$ and $\tilde X$, to sub-packets $X_1, X_2, \tilde X_1, \tilde X_2$ with $X=[X_1, X_2]$, $\tilde X=[\tilde X_1 \tilde X_2],$ and fills in the caches with different sub-packets having  no knowledge, or probably a rough estimation, of the cache nodes' data requests during the HT period (see Fig. 2a). Assuming the packets $X$ and $\tilde X$ to be of length $2L$, the sub-packets $X_1, X_2, \tilde X_1, \tilde X_2$ are of length $L$. In this way, during the LT period, the server sends separate signals $[X_1 \tilde X_1]$ and $[X_2 \tilde X_2]$, each of length $2L,$ to $\text{C}_1$ and $\text{C}_2$, respectively. 

During the HT period (say, 9 PM), the server serves the cache nodes based on their instantaneous data requests. Let $\bigoplus$ be the superposition operator. Also, considering the worst-case scenario in terms of backhaul traffic, assume that the caches $\text{C}_1$ and $\text{C}_2$ request for different packets $X$ and $\tilde X$, respectively. Then, as demonstrated in Fig. 2b, the server broadcasts a single sub-packet $X_2\bigoplus \tilde X_1$ of length $L$. Also, using the accumulated signals, each cache node may use different methods to decode its message of interest (see Section III).  In this way,  as shown in \cite{6763007,6807823}, coded-caching reduces the peak HT backhaul traffic by $50\%$ because, unlike uncoded caching, only a single sub-packet is broadcasted at, say, 9 PM. Finally, it is interesting to note that the presented coded-caching approach is a specific combination of the orthogonal multiple access (OMA) and NOMA (N:non) schemes used in different time slots with proper packet partitioning and signal decoding at the receivers. 

Let us denote the server-cache $i$ channel coefficient by $h_i, i=1,2,$ and define the channel gains as $g_i=|h_i|^2, i=1,2.$ We consider Rayleigh-fading conditions with channel probability density functions (PDFs) $f_{g_i}(u)=\lambda_ie^{-\lambda_iu}, i=1,2,$ where $\lambda_i, i=1,2,$ depends on the long-term channel quality.  Then, the signals received by $\text{C}_1$ and $\text{C}_2$ during the LT period are given by 
\begin{align}\label{eq:eqLT}
    \left\{\begin{matrix}
\left[Y_{\text{C}_1}^\text{LT}(t)\, \tilde Y_{\text{C}_1}^\text{LT}(t)\right]=\sqrt{P}h_1^\text{LT}\left[X_1(t) \tilde X_1(t)\right]+\left[Z_{\text{C}_1,1}(t)\, Z_{\text{C}_1,2}(t)\right], t=1,\ldots,L
\\ 
\left[Y_{\text{C}_2}^\text{LT}(t)\, \tilde Y_{\text{C}_2}^\text{LT}(t)\right]=
\sqrt{P}h_2^\text{LT}\left[X_2(t) \tilde X_2(t)\right]+\left[Z_{\text{C}_2,1}(t)\, Z_{\text{C}_2,2}(t)\right], t=1,\ldots,L
\end{matrix}\right.
\end{align}
while, at the HT period, the received signals are \begin{align}\label{eq:eqHT}
   \left\{\begin{matrix}
Y_{C_1}^\text{HT}(t)=\sqrt{P}h_{1}^\text{HT}S(t)+Z_{\text{C}_1,1}(t), t=1,\ldots,L
\\ 
Y_{C_2}^\text{HT}(t)=\sqrt{P}h_{2}^\text{HT}S(t)+Z_{\text{C}_2,1}(t), t=1,\ldots,L
\\
S(t)=\alpha X_{2}(t)+\sqrt{1-\alpha^2}\tilde X_{1}(t).\,\,\,\,\,\,\,\,\,\,\,\,\,\,\,\,\,\,\,\,\,\,\,\,\,\,\,\,\,\,\,
\end{matrix}\right.
\end{align}
Here, $h_i^\text{LT}$ and $h_i^\text{HT}, i=1,2,$ represent the channel realizations of $h_i$ at LT and HT periods, respectively. Then, $Z_{\text{C}_i,j},i,j=1,2,$ denotes the unit-variance additive Gaussian noise, and $P$ is the server maximum transmit power. Also, $X_i, \tilde X_i, i=1,2,$ are the unit-variance signals of the sub-packets, $S(t)$ represents the unit-variance superimposed signals of $X_2$ and $\tilde X_1$, and $\alpha\in [0,1]$ gives the power partitioning between these signals.

\vspace{-4mm}
\section{Analytical results}
Let us denote the number of information bits in the packets $X$ and $\tilde X$ by $K$ and $\tilde K$, respectively. Also, the information nats are divided between the sub-packets according to 
\begin{align}\label{eq:eqbitpart}
\left\{\begin{matrix}
K=K_1+K_2
\\ 
\tilde K=\tilde K_1+\tilde K_2,
\end{matrix}\right.    
\end{align}
which, defining the code rates $R=\frac{K}{2L}$ and $\tilde R=\frac{\tilde K}{2L},$ leads to \begin{align}\label{eq:eqratepart}
\left\{\begin{matrix}
2R=R_1+R_2
\\ 
2\tilde R=\tilde R_1+\tilde R_2.
\end{matrix}\right.    
\end{align}
Our goal is to design proper rate allocation between sub-packets, i.e., $R_i,\tilde R_i, i=1,2,$ and power split parameter $\alpha$ in (\ref{eq:eqHT}) such that the network STP is maximized. STP is defined as 
\begin{align}\label{eq:eqSTPdef}
\Pr(\text{ST})=\frac{1}{2}\left(\Pr\left(\text{C}_1 \,\text{successful}\right)+\Pr\left(\text{C}_2 \,\text{successful}\right)\right),
\end{align}
i.e., the probability of the event that both cache nodes can decode their intended packets correctly. Depending on the buffering and the coding schemes of the cache nodes, the network may lead to different STPs as follows.
\vspace{-3mm}
\subsection{Joint Decoding at HT Period using Successive Interference Cancellation}
In one  approach, referred to as Method 1 in the following, the cache nodes buffer the signals received in the LT period, and use both maximum ratio combining (MRC) and successive interference cancellation (SIC) for message decoding at HT periods. Let us concentrate on cache node $\text{C}_1$. Receiving $Y_{\text{C}_1}^\text{HT}$ in (\ref{eq:eqHT}) during the HT period and with $\tilde Y_{\text{C}_1}^\text{LT}$ in (\ref{eq:eqLT}) already buffered, $\text{C}_1$ first uses MRC to decode $\tilde X_1(t)$. Then, with a successful decoding of $\tilde X_1(t)$, node $\text{C}_1$ uses the SIC concept to remove $\tilde X_1(t)$ from $Y_{\text{C}_1}^\text{HT}(t)$, leading to an interference-free signal
\vspace{-2mm}
\begin{align}\label{eq:Eqsic1}
\mathcal{Y}_{\text{C}_1}^\text{HT}=\sqrt{P}h_1^\text{HT}\alpha X_2(t)+Z_{\text{C}_1,1}(t).
\end{align}
Finally, the decoder generates the concatenated signal $[Y_{\text{C}_1}^\text{LT}(t) \,\mathcal{Y}_{\text{C}_1}^\text{HT}(t)]$, with $Y_{\text{C}_1}^\text{LT}(t)$ and $\mathcal{Y}_{\text{C}_1}^\text{HT}(t)$ given in (\ref{eq:eqLT}) and (\ref{eq:Eqsic1}), respectively, and decodes the packet $X$ in \emph{one-shot.} In this way, and considering the same procedure in $\text{C}_2$ by using (\ref{eq:eqLT})-(\ref{eq:eqHT}), STP in (\ref{eq:eqSTPdef}) is given by 
\begin{align}\label{eq:eqSTPdef1}
&\Pr(\text{ST})=\frac{1}{2}(\eta_1\gamma_1+\eta_2\gamma_2),\nonumber\\&
\eta_1=\Pr\left(\log\left(1+\frac{(1-\alpha^2)Pg_1^\text{HT}}{1+\alpha^2Pg_1^\text{HT}}+Pg_1^\text{LT}\right)\ge \tilde R_1\right),
\nonumber\\&\gamma_1=\Pr\left(\log\left(1+Pg_1^\text{LT}\right)+\log\left(1+\alpha^2Pg_1^\text{HT}\right)\ge 2R\right)\nonumber\\&
\eta_2=\Pr\left(\log\left(1+\frac{\alpha^2Pg_2^\text{HT}}{1+(1-\alpha^2)Pg_2^\text{HT}}+Pg_2^\text{LT}\right)\ge  R_2\right),\nonumber\\
&\gamma_2=\Pr\left(\log\left(1+Pg_2^\text{LT}\right)+\log\left(1+(1-\alpha^2)Pg_2^\text{HT}\right)\ge 2\tilde R\right).
\end{align}
Here, $\eta_1$ (resp. $\eta_2$) is the probability of successful decoding of $\tilde X_1(t)$ (resp. $X_2(t)$) at $\text{C}_1$ (resp. $\text{C}_2$) using MRC. Then, $\gamma_1$ and $\gamma_2$ give the probability that, removing the interference from the received signal in HT period, the caches can decode their intended signals correctly. Note that in $\gamma_1$ and $\gamma_2$ we have used the results on the maximum achievable rates of parallel Gaussian channels.

Considering Rayleigh-fading conditions, we have 
\begin{align}\label{eq:eqeta11}
&\eta_1=1-\Pr\left(\frac{(1-\alpha^2)Pg_1^\text{HT}}{1+\alpha^2Pg_1^\text{HT}}+Pg_1^\text{LT}\le e^{\tilde R_1}-1\right)=
\nonumber\\&
1-\int_0^{\frac{e^{\tilde R_1}-1}{P}}f_{g_1}(y)\Pr\left(\frac{(1-\alpha^2)Pg_1^\text{HT}}{1+\alpha^2Pg_1^\text{HT}} \le {e^{\tilde R_1}-1}-Py \right)\text{d}y
\nonumber\\&
=e^{-\frac{\lambda_1\left({e^{\tilde R_1}-1}\right)}{P}}+
\int_{\frac{\alpha^2 e^{\tilde R_1}-1}{\alpha^2P}}^{\frac{e^{\tilde R_1}-1}{P}}e^{\lambda_1\left(y+\frac{{e^{\tilde R_1}-1}-Py}{\alpha^2P-(1-\alpha^2)P\left({e^{\tilde R_1}-1}-Py\right)}\right)}\text{d}y,
\end{align}
which can be calculated numerically. Also, following the same procedure, we have
\begin{align}\label{eq:Eqeta22}
    &\eta_2=e^{-\frac{\lambda_2\left(e^{R_2}-1\right)}{P}}
    +\int_{\frac{\left(1-\alpha^2\right)e^{R_2}-1}{\left(1-\alpha^2\right)P}}^{\frac{e^{R_2}-1}{P}}{\lambda_2e^{-\lambda_2\left(y+\frac{e^{R_2}-1-Py}{\alpha^2P-\left(1-\alpha^2\right)P\left(e^{R_2}-1-Py\right)}\right)}}\text{d}y.
\end{align}
The terms $\gamma_i,i=1,2,$ on the other hand, do not have closed-form or easy-to-deal integration expressions. Thus, we use the Jensen's inequality $\frac{1}{n}\sum_{i=1}^n\log(1+x_i)\le \log\left(1+\frac{1}{n}\sum_{i=1}^nx_i\right)$ \cite[Eq. (30)]{7445896} and $f_{g_i}(u)=\lambda_ie^{-\lambda_iu}, i=1,2,$ to rephrase $\gamma_i,i=1,2,$ as
\vspace{-2mm}
\begin{align}\label{eq:eqgamma1}
\gamma_1&\le 1-\Pr\left(g_1^\text{LT}+\alpha^2g_1^\text{HT}\le \frac{2}{P}\left(e^R-1\right)\right)   \nonumber\\&
=1-\int_0^{\frac{2\left(e^R-1\right)}{P}}{f_{g_1}(x)\Pr\left(g_1^\text{HT}\le\frac{\frac{2\left(e^R-1\right)}{P}-x}{\alpha^2}\right)}\text{d}x\nonumber\\&
=e^{\frac{-2\lambda_1\theta}{P}}+\frac{\alpha^2 e^{\frac{-2\lambda_1\theta}{P\alpha^2}}}{\alpha^2-1}\left(1-e^{\frac{-2\lambda_1\left(\alpha^2-1\right)\theta}{P\alpha^2}}\right),
\end{align}
\vspace{-2mm}
\begin{align}\label{eq:eqgamma2}
\gamma_2&\le e^{\frac{-2\lambda_2\theta}{P}}+\frac{\left(\alpha^2-1\right) e^{\frac{-2\lambda_2\theta}{P\left(1-\alpha^2\right)}}}{\alpha^2}\left(1-e^{\frac{-2\lambda_2\alpha^2\theta}{P\left(\alpha^2-1\right)}}\right),
\end{align}
where (\ref{eq:eqgamma2}) follows the same procedure as in (\ref{eq:eqgamma1}). In this way, the optimal rate/power allocation maximizing STP is given by
\begin{align}\label{eq:eqoptprob}
    &\max \frac{1}{2}\{\eta_1\gamma_1+\eta_2\gamma_2\}\nonumber\\
    &\text{s.t. } \,\alpha\in[0,1], R_1+R_2=2R, \,\tilde R_1+\tilde R_2=2\tilde R,
\end{align}
which can be effectively solved by, e.g., exhaustive search or the machine-learning based scheme of \cite{8520925}. 
\vspace{-2mm}
\subsection{Joint Decoding at HT Period without SIC}
Implementation of MRC and SIC, to decode and remove the interference, increases the decoding complexity/delay. Also, SIC suffers from error propagation problem, e.g., \cite{nomaharq}. For these reasons, in Method 2, each cache node decodes its intended packets in one-shot by considering the interference as an additive noise. Here, (\ref{eq:eqSTPdef1}) is rephrased as
\begin{align}\label{eq:eqSTPdef3}
&\Pr(\text{ST})=\frac{1}{2}(\bar \gamma_1+\bar \gamma_2),\nonumber\\&
\bar\gamma_1=\Pr\left(\log\left(1+Pg_1^\text{LT}\right)+\log\left(1+\frac{\alpha^2Pg_1^\text{HT}}{1+(1-\alpha^2)Pg_1^\text{HT}}\right)\ge 2R\right)\nonumber\\&
\bar\gamma_2=\Pr\left(\log\left(1+Pg_2^\text{LT}\right)+\log\left(1+\frac{(1-\alpha^2)Pg_2^\text{HT}}{1+\alpha^2Pg_2^\text{HT}}\right)\ge 2\tilde R\right),
\end{align}
where, using the Jensen's inequality and the same procedure as in (\ref{eq:eqgamma1}), we have
\begin{align}\label{eq:eqjensen2}
&\bar\gamma_1\le e^{-\frac{2\lambda_1\left(e^R-1\right)}{P}}+\int_{\frac{2\left(e^R-1\right)-\frac{\alpha^2}{1-\alpha^2}}{P}}^{\frac{2\left(e^R-1\right)}{P}}\lambda_1e^{-\lambda_1\left(x+\frac{2\left(e^R-1\right)-Px}{\alpha^2P+(1-\alpha^2)P(Px-2\left(e^R-1\right))}\right)}\text{d}x,\nonumber\\&
\bar\gamma_2\le e^{-\frac{2\lambda_2\left(e^{\tilde R}-1\right)}{P}}
+\int_{\frac{2\left(e^{\tilde R}-1\right)-\frac{1-\alpha^2}{\alpha^2}}{P}}^{\frac{2\left(e^{\tilde R}-1\right)}{P}}\lambda_2e^{-\lambda_2\left(x+\frac{2\left(e^{\tilde R}-1\right)-Px}{(1-\alpha^2)P+\alpha^2P(Px-2\left(e^{\tilde R}-1\right))}\right)}\text{d}x,
\end{align}
which can be calculated numerically. 

Finally, note that, replacing (\ref{eq:eqSTPdef3}) into (\ref{eq:eqoptprob}), the optimal performance of the cache nodes in Method 2 is independent of the rate split between the sub-packets. This, although Method 1 gives the best performance in terms of the worst-case peak traffic,
may give an advantage to Method 2, compared to Method 1. This is because in Method 1 the rate split is performed by considering the worst-case condition with the cache nodes requesting for different signals during HT period. However, if the caches request for the same signals during HT period, the rate split scheme of Method 1 is not necessarily optimal. As opposed, in Method 2, the rate split is independent of the caches requested signals in HT periods.

\subsection{Separate Decoding using SIC}
In Methods 1-2, one needs to follow the coding schemes of incremental redundancy hybrid automatic repeat request (HARQ)-based protocols or Raptor codes, e.g., \cite{6164088}, where the message is decoded by concatenating different sub-packets.  Alternatively, in Method 3, we consider the case where, while MRC and SIC are used to decode and remove the interference signal, respectively, each cache node decodes its sub-packets of interest separately. That is, considering $\text{C}_1,$ $X_1$ (resp. $X_2$) is decoded during the LT (resp. HT) period. In this case, the STP (\ref{eq:eqSTPdef1}) is changed to 
\vspace{-1mm}
\begin{align}\label{eq:eqSTPdef2}
&\Pr\left(\text{ST}\right)=\frac{1}{2}\left(\eta_1\breve{\gamma}_{11}\breve{\gamma}_{12}+\eta_2\breve{\gamma}_{21}\breve{\gamma}_{22}\right),\nonumber\\&
\breve{\gamma}_{11}=\Pr\left(\log\left(1+Pg_1^\text{LT}\right)\ge R_1\right)=e^{-\frac{\lambda_1\left(e^{R_1}-1\right)}{P}},\nonumber\\&
\breve{\gamma}_{12}=\Pr\left(\log\left(1+\alpha^2Pg_1^\text{HT}\right)\ge R_2\right)=e^{-\frac{\lambda_1\left(e^{R_2}-1\right)}{\alpha^2P}}
\nonumber\\&
\breve{\gamma}_{21}=\Pr\left(\log\left(1+(1-\alpha^2)Pg_2^\text{HT}\right)\ge \tilde R_1\right)=e^{-\frac{\lambda_2\left(e^{\tilde R_1}-1\right)}{(1-\alpha^2)P}},\nonumber\\&
\breve{\gamma}_{22}=\Pr\left(\log\left(1+Pg_2^\text{LT}\right)\ge \tilde R_2\right)=e^{-\frac{\lambda_2\left(e^{\tilde R_2}-1\right)}{P}},
\end{align}
with $\eta_i,i=1,2,$ given in (\ref{eq:eqSTPdef1}), and (\ref{eq:eqoptprob}) is adapted correspondingly. In (\ref{eq:eqSTPdef2}), $\breve{\gamma}_{11}$ is the probability that  $\text{C}_1$ decodes $X_1$ during the LT period. Also,  $\breve{\gamma}_{12}$ gives the probability that, after decoding and removing $\tilde X_1$, the cache node $\text{C}_1$ correctly decodes $X_2$ in the HT period. Also, the same arguments hold for $\breve{\gamma}_{2i},i=1,2.$

Note that, although Method 1 maximizes the achievable rate/STP, Method 3 has a number of advantages including:
\begin{itemize}
    \item \textbf{Low decoding complexity}: Because, as opposed to Methods 1-2 decoding long codewords of length $2L$, Method 3 is based on decoding sub-packets of length $L$.
    \item \textbf{Efficient HARQ-based transmissions}: In Methods 1-2, all packets are decoded during the HT periods and, in case of decoding failure, the message is retransmitted at that period. Such HARQ-based retransmissions increase the backhauling load at HT period. As opposed, in Method 3, the decoding of the first received sub-packets and all their required HARQ-based retransmissions are performed during the LT period, which reduces the backhauling cost of HARQ.
\end{itemize}
Finally, depending on the considered method, the buffering scheme of the caches during the LT period may change. Particularly, in Methods 1-2 the caches buffer the signals received during LT period without decoding. In Method 3, however, the caches buffer the  sub-packets successfully decoded during LT period. 
\vspace{-1mm}
\subsection{Separate Decoding without SIC}
To further reduce the complexity of Method 3, one can consider the case where, while decoding the sub-packets separately, the cache nodes consider the interference as an additive noise (Method 4). In this case, where the sub-packets are decoded in different LT and HT periods without SIC, the STP is given by
\begin{align}\label{eq:eqSTPdef4}
&\Pr\left(\text{ST}\right)=\frac{1}{2}(\breve{\gamma}_{11}\hat{\gamma}_{12}+\hat{\gamma}_{21}\breve{\gamma}_{22}),\nonumber\\&
\hat{\gamma}_{12}=\Pr\left(\log\left(1+\frac{\alpha^2Pg_1^\text{HT}}{1+(1-\alpha^2)Pg_1^\text{HT}}\right)\ge R_2\right)\nonumber\\&\,\,\,\,\,\,\,\,=\left\{\begin{matrix}
0\,\,\,\,\,\,\,\,\,\,\,\,\,\,\,\,\,\,\,\,\,\,\,\,\,\,\,\,\,\,\,\,\,\,\,\,\,\,\,\,\,\,\,\,\,\,\,\,\,\,\, & \text{if } R_2\ge -\log\left(1-\alpha^2 \right )\\ 
e^{-\lambda_1\frac{\left(e^{R_2}-1 \right )}{\alpha^2P-(1-\alpha^2)P\left(e^{R_2}-1 \right )}} & \text{otherwise}
\end{matrix}\right.
\nonumber\\&
\hat{\gamma}_{21}=\Pr\left(\log\left(1+\frac{(1-\alpha^2)Pg_2^\text{HT}}{1+\alpha^2Pg_2^\text{HT}}\right)\ge \tilde R_1\right)\nonumber\\&=\left\{\begin{matrix}
0\,\,\,\,\,\,\,\,\,\,\,\,\,\,\,\,\,\,\,\,\,\,\,\,\,\,\,\,\,\,\,\,\,\,\,\,\,\,\,\,\,\,\,\,\,\,\,\,\,\,\,\,\, & \text{if } \tilde R_1\ge -2\log\alpha\\ 
e^{-\lambda_2\frac{\left(e^{\tilde R_1}-1 \right )}{(1-\alpha^2)P-\alpha^2P\left(e^{\tilde R_1}-1 \right )}} & \text{otherwise}
\end{matrix}\right.
\end{align}
with $\breve{\gamma}_{11}$ and $\breve{\gamma}_{22}$ given in (\ref{eq:eqSTPdef2}). Also, in (\ref{eq:eqSTPdef4}) we use Rayleigh channel PDFs and some manipulations to derive the probabilities. For further comparisons between Methods 1-4, see Section IV.

\section{Simulation Results}
The simulation results are presented for the cases with $\lambda_1=1$ and $\lambda_2=0.1$, i.e., with $10$ dB difference between the channel gains of the server-cache links, and we define the transmission signal-to-noise ratio (SNR) as $10\log_{10} P$, considering the  additive noises to be unit-variance. Note that we have evaluated the results for different parameter settings, and they show the same qualitative conclusions as those presented in the following. In Figs. 3-4, the results are obtained by optimizing the rate and power allocation. Here, both exhaustive search and the genetic-algorithm based scheme of \cite{8520925} have been used which have ended up in the same results, indicating the accuracy of the optimization process. In Fig. 5, we study the effect of rate/power allocation.
\begin{figure}
\vspace{-3mm}
\centering
  \includegraphics[width=0.6\columnwidth]{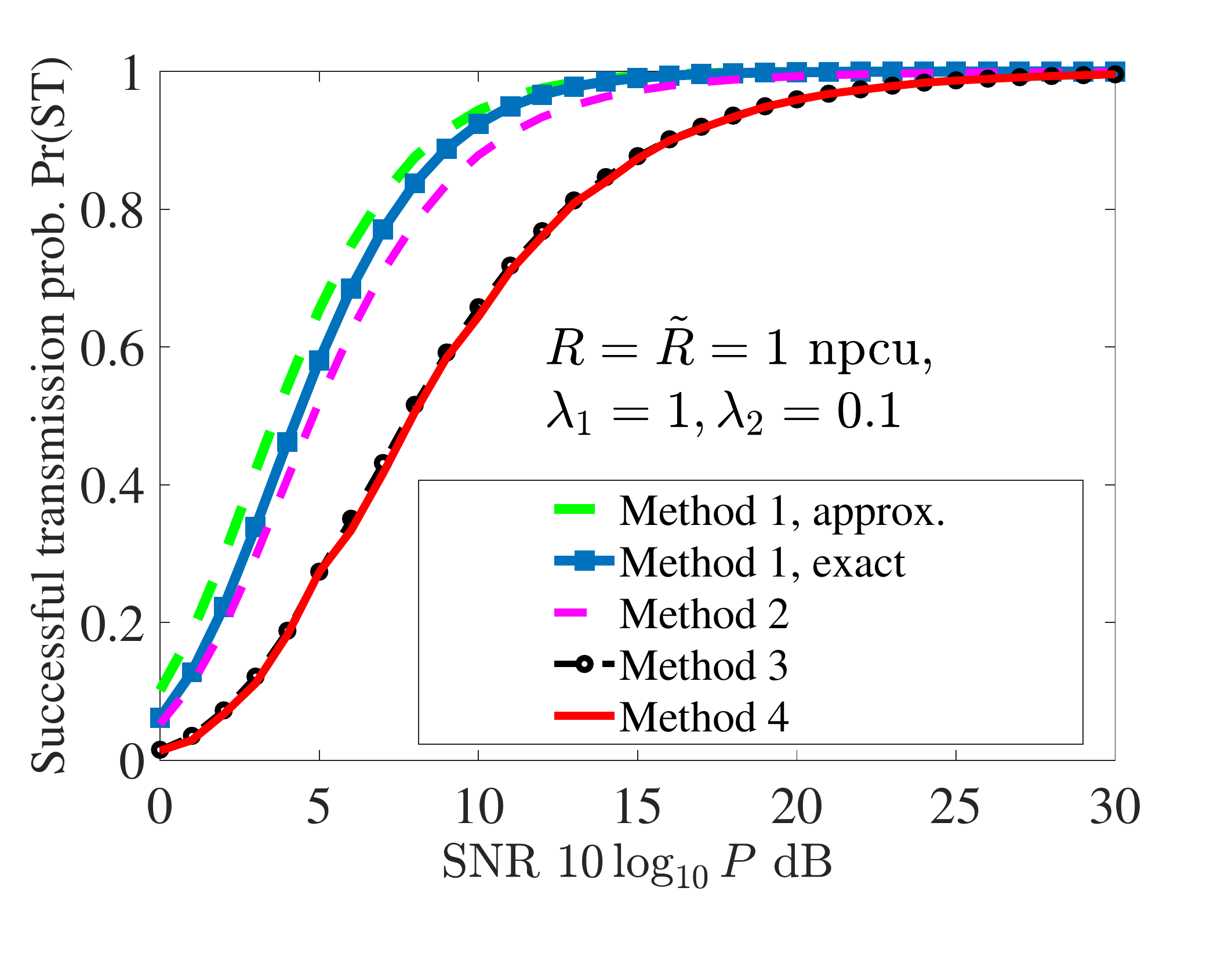}\\\vspace{-3mm}
\caption{Comparison between the STP of Methods 1-4, $\lambda_1=1,$ $\lambda_2=0.1,$ $R=\tilde R=1$ npcu. The results are obtained by optimal rate and power allocation.}
\label{fig:fig_ivd9}
\vspace{-2mm}
\end{figure}

Considering $R=\tilde R=1$ nats-per-channel-use (npcu), Fig. 3 compares the performance of Methods 1-4. Also, the figure verifies the tightness of the Jensen's inequality-based approximation results of (\ref{eq:eqgamma1})-(\ref{eq:eqgamma2}). Then, Fig. 4 shows the STP versus the data rates $R=\tilde R$ for the cases with different decoding/buffering methods and transmission SNRs.

Considering Methods 1 and 3, with joint and separate decoding on the sub-packets, respectively, Fig. 5 studies the effect of optimal rate and power allocation on the network STP. Particularly, the figure compares the optimal results obtained by (\ref{eq:eqoptprob}) with the cases using uniform power allocation, i.e., with $\alpha=\frac{\sqrt{2}}{2}$ in (\ref{eq:eqHT}), and/or equal rate split, i.e., $R_i=\tilde R_i, i=1,2$. According to the figures, the following conclusions can be drawn:

\begin{figure}
\vspace{-3mm}
\centering
  \includegraphics[width=0.6\columnwidth]{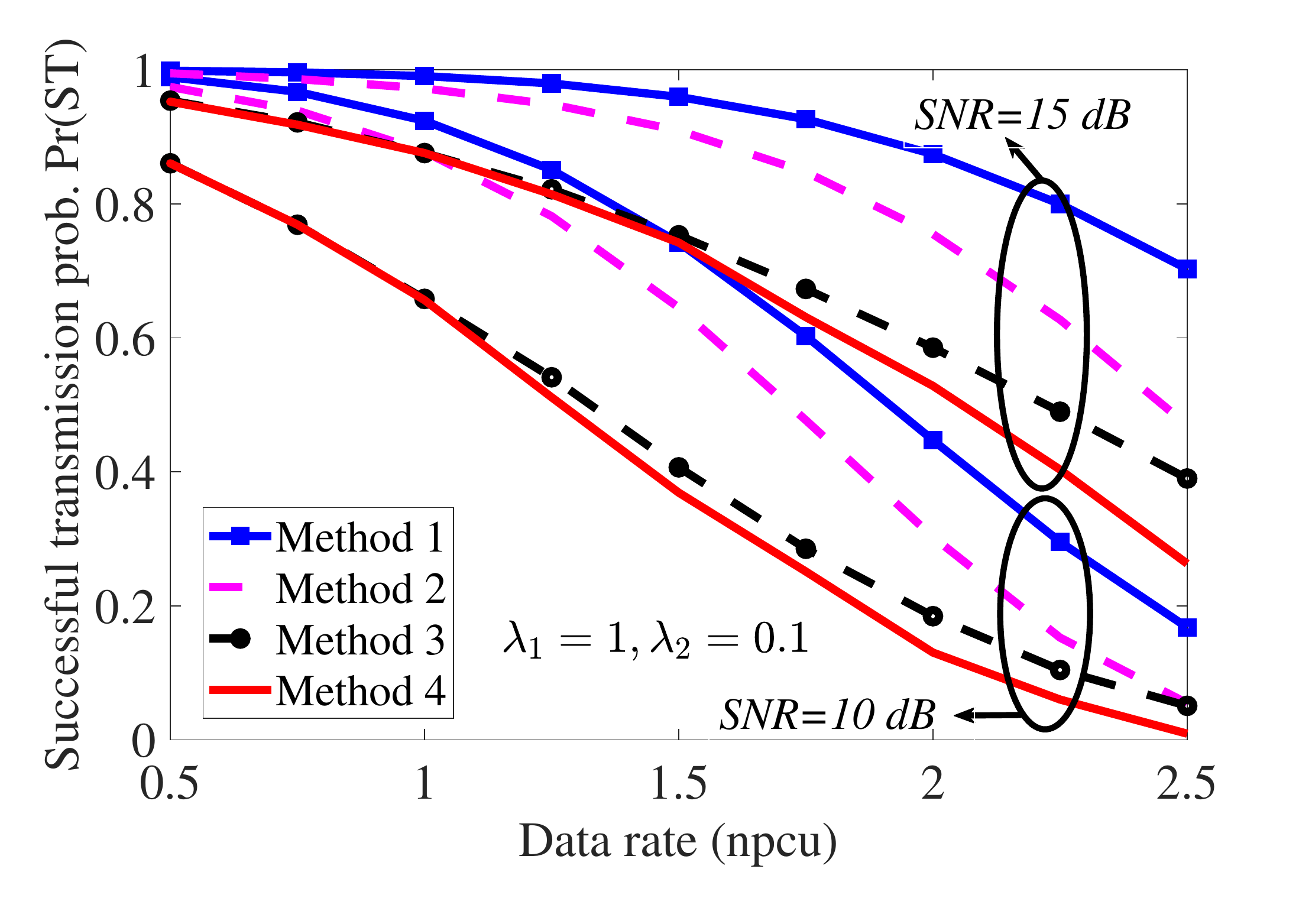}\\\vspace{-2mm}
\caption{Achievable STP versus the data rate for different methods. The results are obtained by optimal rate and power allocation and $\lambda_1=1, \lambda_2=0.1.$}
\label{fig:fig_ivd10}
\vspace{-0mm}
\end{figure}

\begin{figure}
\vspace{-3mm}
\centering
  \includegraphics[width=0.6\columnwidth]{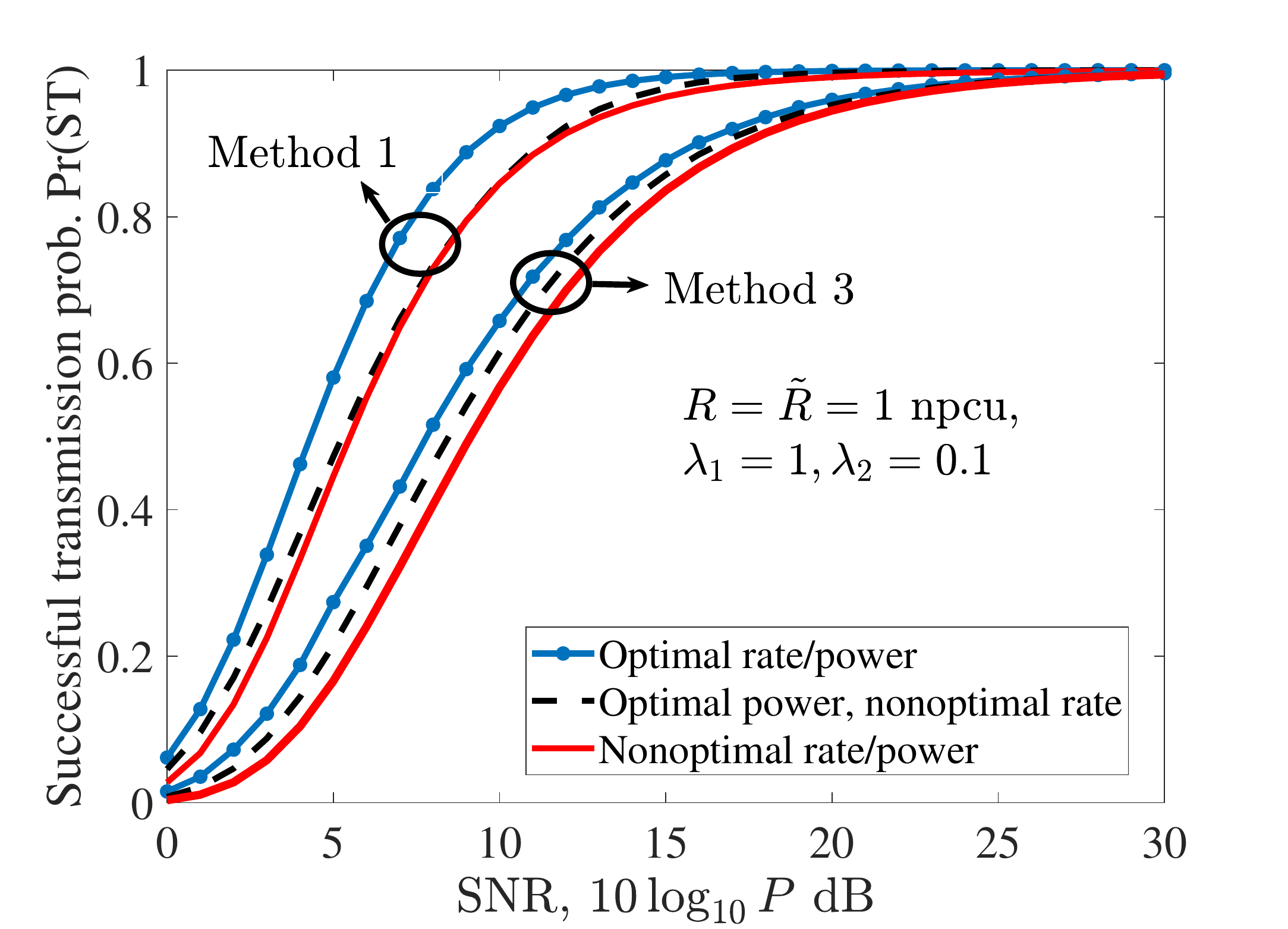}\\\vspace{-2mm}
\caption{On the effect of optimal rate and power allocation in Methods 1 and 3, $R=\tilde R=1$ npcu and $\lambda_1=1, \lambda_2=0.1$.}
\label{fig:fig_ivd11}
\vspace{-0mm}
\end{figure}

\begin{itemize}
  \item The approximation results of (\ref{eq:eqgamma1})-(\ref{eq:eqgamma2}) properly approximate the  probabilities $\gamma_i, i=1,2$ (Fig. 3. Also, the same point is observed for $\bar \gamma_i,i=1,2,$ in (\ref{eq:eqjensen2}) although not shown in the figure).  Thus, the approximations can be well utilized for the  performance evaluation of Methods 1-2, i.e., in the cases with joint decoding of the sub-packets.
  \item Compared to the cases with separate decoding of sub-packets, i.e., Methods 3-4, considerable STP improvement is observed by joint decoding of the sub-packet, i.e., Methods 1-2  (Figs. 3-4). However, as explained in Section III.C, the STP increment of Methods 1-2 is at the cost of decoding complexity and possible HARQ-based retransmissions at HT periods. On the other hand, for both cases with joint and separate decoding of the sub-packets, using SIC-based  interference cancellation leads to marginal performance improvement at low rates while its effect increases with the data rate (Fig. 4). 
  Finally, as the data rate increases, the performance gap between Methods 2 and 3 decreases, i.e., one can reach the same STP as in the cases with joint sub-packet decoding of interference-affected signals by separate sub-packets decoding if the interference signals are removed using SIC. (Fig. 4).
  \item For all parameter settings, Method 1 leads to the highest STP, compared to Methods 2-4, at the cost of decoding delay/possible retransmissions at HT periods (Figs. 3-4). For instance, with the parameter settings of Fig. 4 and data rate 1.5 npcu, the implementation of Method 1 with transmit SNR 10 dB results in the same STP, $80\%$, as in the cases with Method 4 and SNR 15 dB, i.e., using advanced decoding methods leads to 5 dB gain in SNR (Fig. 4). 
  \item For both cases with and without interference cancellation (Methods 1 and 3), optimal rate allocation leads to considerable STP increment (Fig. 5. The same conclusion is observed in Methods 2 and 4, although not presented in the figure). Also, the relative performance gain of optimal rate split increases in the cases with interference cancellation. Finally, considering the interference as additive noise, optimal power allocation between sub-packets during HT period increases the STP. However, with interference cancellation and joint decoding of sub-packets, the effect of optimal power allocation between HT period sub-packets is marginal (Fig. 5).
\end{itemize}

\section{Conclusions}
This paper studied the performance coded-caching networks in the cases with adaptive rate/power allocation and different decoding/buffering schemes. As we showed, joint decoding of the sub-packets at HT periods leads to considerable performance improvement of coded-caching setups. Also, for different decoding schemes, optimal rate split between the sub-packets increases the STP considerably while optimal power allocation between the sub-packets of HT period only improves the STP if SIC-based receiver is not implemented and the sub-packets are decoded separately.


\vspace{-2mm}
\bibliographystyle{IEEEtran} 
\bibliography{main.bib}
\vfill
\end{document}